# Fractional Programming for Communication Systems—Part II: Uplink Scheduling via Matching

Kaiming Shen, *Student Member, IEEE*, and Wei Yu, *Fellow, IEEE*

*Abstract*—This two-part paper develops novel methodologies for using fractional programming (FP) techniques to design and optimize communication systems. Part I of this paper proposes a new quadratic transform for FP and treats its application for continuous optimization problems. In this Part II of the paper, we study discrete problems, such as those involving user scheduling, which are considerably more difficult to solve. Unlike the continuous problems, discrete or mixed discrete-continuous problems normally cannot be recast as convex problems. In contrast to the common heuristic of relaxing the discrete variables, this work reformulates the original problem in an FP form amenable to distributed combinatorial optimization. The paper illustrates this methodology by tackling the important and challenging problem of uplink coordinated multi-cell user scheduling in wireless cellular systems. Uplink scheduling is more challenging than downlink scheduling, because uplink user scheduling decisions significantly affect the interference pattern in nearby cells. Further, the discrete scheduling variable needs to be optimized jointly with continuous variables such as transmit power levels and beamformers. The main idea of the proposed FP approach is to decouple the interaction among the interfering links, thereby permitting a distributed and joint optimization of the discrete and continuous variables with provable convergence. The paper shows that the well-known weighted minimum mean-square-error (WMMSE) algorithm can also be derived from a particular use of FP; but our proposed FP-based method significantly outperforms WMMSE when discrete user scheduling variables are involved, both in term of run-time efficiency and optimizing results.

*Index Terms*—Fractional programming (FP), Lagrangian dual transform, user scheduling, discrete power control, discrete beamforming

## I. Overview

**F**RACTIONAL programming (FP) is a valuable tool for the design and optimization of communication systems, because of the prominent role fractional terms—in particular the signal-to-interference-plus-noise (SINR) ratio—plays in the performance analysis of communication links. Part I of this paper [3] proposes a novel *quadratic transform* technique to tackle FP problems involving multiple ratios, which are frequently encountered in communication system design, but are typically beyond the capabilities of classic FP techniques, such as Schaible's transform [4] and Dinkelbach's method

Manuscript submitted to IEEE Transactions on Signal Processing January 9, 2017; revised August 31 and December 9, 2017; accepted February 13, 2018. This work is supported in part by Natural Science and Engineering Research Council (NSERC) and in part by Huawei Technologies Canada. The materials in this paper have been presented in part in Asilomar Conference on Signals, Systems, and Computers, Nov. 2015, Pacific Grove, CA [1] and in IEEE International Conference on Acoustic, Speech, and Signal Processing (ICASSP), May 2016, Shanghai, China [2]. The authors are with the Edward S. Rogers Sr. Department of Electrical and Computer Engineering, University of Toronto, Toronto, ON M5S 3G4, Canada (e-mails: {kshen,weiyu}@ece.utoronto.ca).

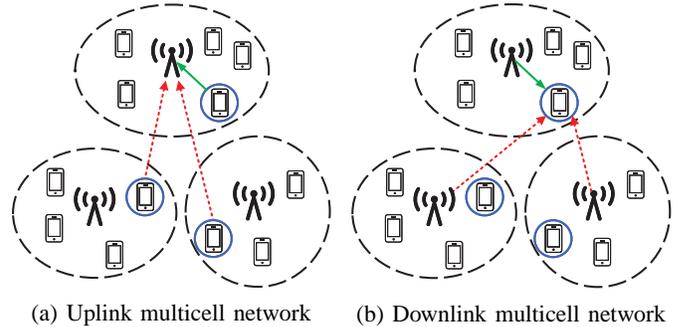

(a) Uplink multicell network    (b) Downlink multicell network

Fig. 1. Interference pattern depends on the user scheduling in the neighboring cells in the uplink, but not so in the downlink. Here, the solid lines represent the desired signal; the dashed lines represent the interfering signal; the scheduled user terminal in each cell is circled.

[5]. It is shown in Part I that the quadratic transform can greatly facilitate the power control, beamforming, and energy efficiency maximizations.

The Part II of this paper explores the use of FP for optimization problems that involve discrete variables within the $\log(1+\text{SINR})$ rate expressions—in particular the problem of coordinated multi-cell uplink user scheduling in wireless cellular networks, where the optimization parameters are the selection of which users to schedule in each cell, along with their power and beamforming vectors. The scheduling problem in the uplink is more challenging than in the downlink, because the uplink interference pattern depends strongly on the scheduling decisions of the neighboring cells, whereas in the downlink, the interference pattern does not depend on scheduling decisions, as illustrated in Fig. 1.

There is a fundamental difference between the uplink scheduling problem and the various continuous FP problems treated in Part I. Due to the discrete variables involved in scheduling, the quadratic transform, which is used extensively in Part I to transform the problem into a sequence of convex problems, is no longer sufficient by itself.

Discrete optimization problems are traditionally tackled using the common heuristic of relaxing the discrete constraints into continuous ones, then quantizing the solution after solving the relaxed problem. The difficulty with this traditional approach is that the resulting relaxed problem is not necessarily always easy to solve, and the final quantization step may not be easy to design (naive rounding scheme is usually suboptimal).

A key observation of this paper is that instead of relaxing the discrete scheduling variables and trying to convexify the problem, we can take advantage of the fact that specific class of discrete optimization problems, namely *the weighted bipartite matching problem*, can be efficiently solved in polynomial-time using established methods such as the Hungarian algorithm [6] and the auction algorithm [7]. By recasting the uplink scheduling problem in a weighted bipartite matching form, via a proposed new technique named *Lagrangian dual transform*, which can "move" the fractional SINR term to the outside of the logarithm, and subsequently allow the quadratic transform and bipartite matching method to be applied, an overall efficient uplink scheduling algorithm can be designed.

The proposed scheme is markedly different from the existing approaches to the uplink scheduling problem studied extensively in the literature. The uplink scheduling schemes implemented in practice [8], [9] are often based on channel quality alone or assume worst-case interference. Because of the difficulty in quantifying the cross-cell interference, most existing uplink scheduling algorithms are heuristic in nature. For example, [10]–[12] propose various heuristics to approximate the uplink SINR. The game theoretical approaches are considered [13], [14], but not in a rigorous way. Other common heuristics include opportunistic method [10], [15], greedy method [16]–[19], relaxation method [20], and clustering method [21].

While the use of these heuristics is justified by the practical consideration in the cost for obtaining channel state information (CSI), this paper aims to show how much better the performance of uplink scheduling algorithm can be if CSI is available. In this realm, [22] shows that the uplink scheduling and power control problem can be solved globally by a monotonic optimization, but in exponential time. The optimality of uplink scheduling is also considered in [23] under some very specific channel conditions. Moreover, iterative scheduling and power control schemes are proposed in [12], [24], which do not perform as well as the scheme proposed in this paper for the uplink. We remark here that although the scheduling problem can be thought of as a power control problem, the approach of relying of power control for scheduling (thereby sidestepping the difficulty of discrete optimization) typically do not perform well, because it can result in *premature turning-off*, as discussed in Section IV-B.

The main goal of this paper is to show that the FP techniques can be applied to the uplink scheduling problem, and that cooperation across the multiple cells in a wireless cellular network has the potential to significantly improve the overall performance of the network. Toward this end, we make the following contributions:

- *Joint Uplink Scheduling and Power Control:* The objective is to optimally schedule uplink users and to set their transmit power levels jointly across multiple cells so as to maximize the network utility in a single-input single-output (SISO) network. The problem involves mixed continuous variables (power) and discrete variables (uplink scheduling); it is quite challenging, because scheduling and power decisions in each cell significantly affect the interference patterns in neighboring cells. This paper proposes an FP-based reformulation that allows power control and uplink scheduling to be determined jointly and in a distributed fashion with the assistance of some auxiliary variables. We remark that this approach can be further extended to apply to the device-to-device (D2D) [25] and the full-duplex [26] settings.

- *Joint User Scheduling and Beamforming:* The objective is to schedule uplink users and to set their transmit beamformers jointly across multiple cells so as to maximize the network utility in a multiple-input multiple-output (MIMO) network. The key step is to incorporate a further FP reformulation involving vector variables. The resulting reformulation allows the optimization of discrete and continuous variables in a joint and distributed fashion using bipartite matching algorithms. Moreover, when the beamforming variable is also discrete (i.e., the beamforming vector must be selected from a given codebook), we propose a nearest point projection scheme which is more efficient than the direct searching; this scheme works for discrete power control as well.

- *FP versus WMMSE:* The proposed FP framework is compared with the well-known WMMSE algorithm for beamforming. Although originally motivated from a minimum-mean-square-error perspective [27], [28], the WMMSE algorithm can be shown to be closely related to FP. This paper shows however that our proposed way of applying FP to scheduling is more advantageous than WMMSE when dealing with discrete scheduling variables.

The notation follows that in Part I. In particular, denote $\mathbb{R}$ as the set of real numbers, $\mathbb{R}_+$ the set of nonnegative real numbers, and $\mathbb{R}_{++}$ the set of strictly positive real numbers. Denote $\mathbb{C}$ as the set of complex numbers. Denote $\mathbb{S}_{++}$ as the set of symmetric positive definite matrices.

## II. QUADRATIC TRANSFORM

We briefly review the quadratic transform in this section; more details can be found in the Part I of this paper [3].

*Theorem 1 (Quadratic Transform [3]):* Given a nonempty constraint set $\mathcal{X} \subseteq \mathbb{R}^d$, a nonnegative function $A(\mathbf{x})\colon \mathbb{R}^d \to \mathbb{R}_+$, and a positive function $B(\mathbf{x})\colon \mathbb{R}^d \to \mathbb{R}_{++}$, where $d \in \mathbb{N}$, a (single-ratio) FP problem is

$$\underset{\mathbf{x}}{\operatorname{maximize}} \quad \frac{A(\mathbf{x})}{B(\mathbf{x})} \tag{1a}$$
$$\text{subject to} \quad \mathbf{x} \in \mathcal{X}. \tag{1b}$$

This problem is equivalent to

$$\underset{\mathbf{x},\, y}{\operatorname{maximize}} \quad 2y\sqrt{A(\mathbf{x})} - y^2 B(\mathbf{x}) \tag{2a}$$
$$\text{subject to} \quad \mathbf{x} \in \mathcal{X},\ y \in \mathbb{R}. \tag{2b}$$

This quadratic transform works for the multiple-ratio case as stated in the following.

*Corollary 1 (Sum-of-Ratios Problem [3]):* Given $M$ pairs of nonnegative function $A_m(\mathbf{x}) : \mathbb{R}^d \to \mathbb{R}_+$ and positive function $A_m(\mathbf{x}) : \mathbb{R}^d \to \mathbb{R}_{++}$ for $m = 1, \ldots, M$, the *sum-of-ratios* problem

$$\underset{\mathbf{x}}{\operatorname{maximize}} \quad \sum_{m=1}^{M} \frac{A_m(\mathbf{x})}{B_m(\mathbf{x})} \tag{3a}$$
$$\text{subject to} \quad \mathbf{x} \in \mathcal{X} \tag{3b}$$

is equivalent to

$$\underset{\mathbf{x},\mathbf{y}}{\text{maximize}} \quad \sum_{m=1}^{M} \left(2y_m\sqrt{A_m(\mathbf{x})} - y_m^2 B_m(\mathbf{x})\right) \quad (4a)$$
$$\text{subject to} \quad \mathbf{x} \in \mathcal{X} \quad (4b)$$

where $\mathbf{y}$ refers to $(y_1, y_2, \ldots, y_M)$.

The quadratic transform in Theorem 1 can be further extended for the multidimensional and complex problem.

*Theorem 2 (Multidimensional and Complex FP [3]):* Given function $\boldsymbol{\alpha}(\mathbf{x}): \mathbb{C}^{d_1} \to \mathbb{C}^{d_2}$, function $\mathbf{B}(\mathbf{x}): \mathbb{C}^{d_1} \to \mathbb{S}_{++}^{d_2 \times d_2}$ and constraint set $\mathcal{X} \subseteq \mathbb{C}^{d_1}$, where $d_1, d_2 \in \mathbb{N}$, a *multidimensional and complex* FP problem of

$$\underset{\mathbf{x}}{\text{maximize}} \quad \boldsymbol{\alpha}^\dagger(\mathbf{x})(\mathbf{B}(\mathbf{x}))^{-1}\boldsymbol{\alpha}(\mathbf{x}) \quad (5a)$$
$$\text{subject to} \quad \mathbf{x} \in \mathcal{X}. \quad (5b)$$

is equivalent to

$$\underset{\mathbf{x},\mathbf{y}}{\text{maximize}} \quad 2\text{Re}\left\{\mathbf{y}^\dagger \boldsymbol{\alpha}(\mathbf{x})\right\} - \mathbf{y}^\dagger \mathbf{B}(\mathbf{x})\mathbf{y} \quad (6a)$$
$$\text{subject to} \quad \mathbf{x} \in \mathcal{X}, \, \mathbf{y} \in \mathbb{C}^{d_2}. \quad (6b)$$

This multidimensional and complex quadratic transform can also be extended to the multiple-ratio case [3].

## III. LAGRANGIAN DUAL TRANSFORM

The quadratic transform as stated above is the core FP technique used in Part I for treating the continuous problems. When it comes to the discrete problems of user scheduling, we need to introduce a new FP technique named *Lagrangian dual transform*.

### A. Target Problem

Optimization problem for communication system design often involves data rates expressed as logarithmic functions of SINR, i.e., $\log(1 + \text{SINR})$. Part I of this paper [3] proposes two different approaches for applying FP to such problems. In the direct FP, the quadratic transform is immediately applied to the log-function of the ratio to decouple the numerator and denominator, while in the closed-form FP, a Lagrangian dual transform is first applied to take the ratio out of the logarithm. For continuous optimization problems, the two approaches give comparable performance. However, for discrete scheduling problems involving $\log(1 + \text{SINR})$, the second approach of using Lagrangian dual transform becomes indispensable.

This paper develops the Lagrangian dual transform technique that accomplishes the task "moving" SINR to the outside of logarithm. This technique plays a crucial role in addressing the discrete scheduling problems, because it allows a subsequent quadratic transform to express all optimization variables in linear terms. This section gives a detailed derivation of the Lagrangian dual transform technique with a constructive proof of the main result.

The target problem is a weighted sum-of-logarithms maximization[1]:

$$\underset{\mathbf{x}}{\text{maximize}} \quad \sum_{m=1}^{M} w_m \log\left(1 + \frac{A_m(\mathbf{x})}{B_m(\mathbf{x})}\right) \quad (7a)$$
$$\text{subject to} \quad \mathbf{x} \in \mathcal{X} \quad (7b)$$

where $w_m$'s are nonnegative weights, $A_m(\mathbf{x})$'s are nonnegative functions and $B_m$'s are positive functions for all $m$, and $\mathcal{X}$ is a nonempty constraint set. The above formulation is often used to model the weighted sum rate maximization problem of a communication network. The ratio $A_m/B_m$ can be physically interpreted as the SINR term. The problem (7) has no known convex reformulation. Further, the constraint represented by $\mathcal{X}$ is not necessarily compact, i.e., the variable $\mathbf{x}$ may be discrete or mixed discrete-continuous.

### B. Transform

The main result is the following Lagrangian dual transform capable of converting (7) to a sum-of-ratios form.

*Theorem 3 (Lagrangian Dual Transform):* The weighted sum-of-logarithms problem (7) is equivalent to

$$\underset{\mathbf{x},\boldsymbol{\gamma}}{\text{maximize}} \quad f_r(\mathbf{x}, \boldsymbol{\gamma}) \quad (8a)$$
$$\text{subject to} \quad \mathbf{x} \in \mathcal{X} \quad (8b)$$

where $\gamma_m$ is introduced as an auxiliary variable introduced for each ratio term $A_m(\mathbf{x})/B_m(\mathbf{x})$; the new objective function $f_r$ is defined by

$$f_r(\mathbf{x},\boldsymbol{\gamma}) = \sum_{m=1}^{M} w_m \log(1+\gamma_m) - \sum_{m=1}^{M} w_m \gamma_m + \underbrace{\sum_{m=1}^{M} \frac{w_m(1+\gamma_m)A_m(\mathbf{x})}{A_m(\mathbf{x}) + B_m(\mathbf{x})}}_{\text{Sum-of-ratio term}}. \quad (9)$$

The two problems are equivalent in the sense that $\mathbf{x}$ is the solution to (7) if and only if it is the solution to (8), and the optimal objective values of these two problems are also equal.

*Proof:* Observe that $f_r$ is a concave differentiable function over $\boldsymbol{\gamma}$ when $\mathbf{x}$ is held fixed, so $\boldsymbol{\gamma}$ can be optimally determined by setting each $\partial f_r/\partial \gamma_m$ to zero, i.e., $\gamma^\star = A_m(\mathbf{x})/B_m(\mathbf{x})$. Substituting this $\gamma^\star$ back in $f_r$ recovers the weighted sum-of-logarithms objective function in (7a) exactly. The equivalence is therefore established. ∎

Theorem 3 can be extended to the multidimensional and complex case as stated in the following theorem.

*Theorem 4 (Lagrangian Dual Transform in Multidimensional and Complex Case):* Given a sequence of multidimensional and complex functions $\boldsymbol{\alpha}(\mathbf{x}): \mathbb{C}^{d_1} \to \mathbb{C}^{d_2}$ for $m = 1, \ldots, M$, a multidimensional function $\mathbf{B}(\mathbf{x}): \mathbb{C}^{d_1} \to \mathbb{S}_{++}^{d_2 \times d_2}$ and a nonempty constraint set $\mathcal{X} \subseteq \mathbb{C}^{d_1}$, where $d_1, d_2 \in \mathbb{N}$, a multidimensional and complex logarithmic FP problem

$$\underset{\mathbf{x}}{\text{maximize}} \quad \sum_{m=1}^{M} w_m \log\left(1 + \boldsymbol{\alpha}_m^\dagger(\mathbf{x})\mathbf{B}_m^{-1}(\mathbf{x})\boldsymbol{\alpha}_m(\mathbf{x})\right) \quad (10a)$$
$$\text{subject to} \quad \mathbf{x} \in \mathcal{X} \quad (10b)$$

---
[1]For ease of notation, we use the natural logarithm throughout the paper.

can be also recast to the form of (8) where the new objective function $f_r$ is defined in (11) at the bottom of the page.

*Proof:* Since $f_r$ is analytic in the complex plane and also $f_r$ is concave over $\boldsymbol{\gamma}$ for fixed $\mathbf{x}$, we take its complex derivative and solve each $\partial f_r/\partial \gamma_m = 0$. The optimal $\gamma^\star$ is easily seen as $\boldsymbol{\alpha}_m^\dagger(\mathbf{x})\mathbf{B}_m^{-1}(\mathbf{x})\boldsymbol{\alpha}_m(\mathbf{x})$. Substituting this $\gamma^\star$ back in $f_r$ recovers the weighted sum-of-logarithms objective function in (10a) exactly, thereby establishing the equivalence. ∎

### C. Constructive Derivation

To provide insight on how the above transform is obtained, we revisit the weighted sum-of-logarithms problem (7) from a Lagrangian dual perspective, and provide an alternative constructive proof of Theorem 3.

First, by introducing a new variable $\gamma_m$ to replace each ratio term inside the logarithm, (7) can be rewritten as

$$\underset{\mathbf{x}, \boldsymbol{\gamma}}{\text{maximize}} \quad \sum_{m=1}^{M} w_m \log(1 + \gamma_m) \tag{12a}$$

$$\text{subject to} \quad \mathbf{x} \in \mathcal{X} \tag{12b}$$

$$\gamma_m \leq \frac{A_m(\mathbf{x})}{B_m(\mathbf{x})}, \ \forall m = 1, \ldots, M, \tag{12c}$$

where $\boldsymbol{\gamma}$ refers to a collection of auxiliary variables $\{\gamma_1, \ldots, \gamma_M\}$. The above optimization can be thought of as an outer optimization over $\mathbf{x}$ and an inner optimization over $\gamma_m$ with fixed $\mathbf{x}$. The inner optimization is as follows:

$$\underset{\boldsymbol{\gamma}}{\text{maximize}} \quad \sum_{m=1}^{M} w_m \log(1 + \gamma_m) \tag{13a}$$

$$\text{subject to} \quad \gamma_m \leq \frac{A_m(\mathbf{x})}{B_m(\mathbf{x})}, \ \forall m = 1, \ldots, M. \tag{13b}$$

The solution to this inner optimization is obviously that $\gamma_m$ should satisfy (13b) with equality. But, let's express the problem in a different way. Note that (13) is a convex optimization in $\boldsymbol{\gamma}$, so the *strong duality* [29] holds. Introduce the dual variable $\lambda_m$ for each inequality constraint in (13b) and form the Lagrangian function

$$L(\boldsymbol{\gamma}, \boldsymbol{\lambda}) = \sum_{i=1}^{M} w_m \log(1 + \gamma_m) - \sum_{m=1}^{M} \lambda_m \left( \gamma_m - \frac{A_m(\mathbf{x})}{B_m(\mathbf{x})} \right). \tag{14}$$

Due to strong duality, the optimization (13) is equivalent to the dual problem

$$\underset{\boldsymbol{\lambda} \succeq \mathbf{0}}{\text{minimize}} \quad \underset{\boldsymbol{\gamma}}{\text{maximize}} \quad L(\boldsymbol{\gamma}, \boldsymbol{\lambda}). \tag{15}$$

Let $(\boldsymbol{\gamma}^\star, \boldsymbol{\lambda}^\star)$ be the saddle point of the above. It must satisfy the first-order condition $\partial L/\partial \gamma_m = 0$:

$$\lambda_m^\star = \frac{w_m}{1 + \gamma_m^\star}, \ \forall m = 1, \ldots, M. \tag{16}$$

But from the trivial solution to the optimization problem (13), we already know that $\gamma_m^\star = A_m(\mathbf{x})/B_m(\mathbf{x})$, so

$$\lambda_m^\star = \frac{w_m B_m(\mathbf{x})}{A_m(\mathbf{x}) + B_m(\mathbf{x})}, \ \forall m = 1, \ldots, M. \tag{17}$$

Note that $\lambda_m^\star \geq 0$ is automatically satisfied here. Using (17) in (15), problem (13) can then be reformulated as

$$\underset{\boldsymbol{\gamma}}{\text{maximize}} \quad L(\boldsymbol{\gamma}, \boldsymbol{\lambda}^\star). \tag{18}$$

Furthermore, combining with the outer maximization over $\mathbf{x} \in \mathcal{X}$ and after some algebra, we find (18) to be exactly the same as the maximization of (9) in Theorem 3.

We remark that a similar Lagrangian dual procedure based on the multidimensional complex differentiation can be derived for Theorem 4; the details are omitted.

## IV. Joint Uplink Scheduling and Power Control

We now consider the coordinated uplink scheduling and power control problem as an application of FP to discrete optimization.

### A. Problem Formulation

Consider the uplink of a wireless cellular network. Let $\mathcal{B}$ be the set of base-stations (BSs) deployed in the network, and let $\mathcal{K}_i$ be the set of users who are associated with BS $i$. Each BS $i$ together with its associated users in $\mathcal{K}_i$ forms a cell. In every time-slot, users are scheduled for uplink transmission on a cell basis. In this section, the BSs and the users are assumed to be equipped with a single antenna each; extension to the multiple-antenna case involving beamforming optimization is considered in the next section. For the user scheduling and power control purpose, introduce variable $s_i \in \mathcal{K}_i$ to denote the user to be scheduled at BS $i$, and introduce variable $p_k$ to denote the transmit power level of user $k$ if it gets scheduled for uplink transmission. Let $h_{i,k} \in \mathbb{C}$ be the uplink channel coefficient from user $k$ to BS $i$; let $\sigma^2$ be the additive white Gaussian background noise (AWGN) power. Given a set of weights $w_k$ that reflect the user priorities in each time-slot, we have the following weighted sum rate maximization objective:

$$f_o(\mathbf{s}, \mathbf{p}) = \sum_{i \in \mathcal{B}} w_{s_i} \log \left( 1 + \frac{|h_{i,s_i}|^2 p_{s_i}}{\sum_{j \neq i} |h_{i,s_j}|^2 p_{s_j} + \sigma^2} \right). \tag{19}$$

The joint scheduling and power control problem in an uplink SISO network can be written as

$$\underset{\mathbf{s}, \mathbf{p}}{\text{maximize}} \quad f_o(\mathbf{s}, \mathbf{p}) \tag{20a}$$

$$\text{subject to} \quad 0 \leq p_k \leq P_{\max} \tag{20b}$$

$$s_i \in \mathcal{K}_i \cup \{\varnothing\} \tag{20c}$$

where $\mathbf{s}$ denotes the collection of scheduling variables $\{s_i\}_{i \in \mathcal{B}}$, $\mathbf{p}$ denotes the collection of power variables $\{p_k\}_{k \in \bigcup_{i \in \mathcal{B}} \mathcal{K}_i}$,

---

$$f_r(\mathbf{x}, \boldsymbol{\gamma}) = \sum_{m=1}^{M} w_m \log(1 + \gamma_m) - \sum_{m=1}^{M} w_m \gamma_m + \sum_{m=1}^{M} w_m (1 + \gamma_m) \boldsymbol{\alpha}_m^\dagger(\mathbf{x})(\boldsymbol{\alpha}_m(\mathbf{x})\boldsymbol{\alpha}_m^\dagger(\mathbf{x}) + \mathbf{B}_m(\mathbf{x}))^{-1}\boldsymbol{\alpha}_m(\mathbf{x}) \tag{11}$$

$P_{\max}$ is the maximum transmit power level of the user, $\varnothing$ refers to the decision of not scheduling any user. Because of the SISO setting, at most one user can be scheduled in each cell $i$; we set $s_i = k$ if some user $k$ is scheduled in the cell, and set $s_i = \varnothing$ otherwise.

The above problem is difficult to tackle directly due to the fact that the uplink scheduling decisions have significant impact on the interference pattern. A particular scheduling decision $s_i$ in cell $i$ strongly influences the scheduling decisions $s_j$ in its neighboring cells. In addition, even when the discrete variable $\mathbf{s}$ is held fixed, solving for the power variable $\mathbf{p}$ in (20) is still nontrivial, because the objective function is nonconvex.

*B. Implicit Scheduling by Power Control*

Before proceeding to the proposed FP approach, we discuss an alternative perspective of treating the uplink scheduling problem as a power control problem, and explain why the corresponding optimization method would not produce good results numerically.

As opposed to formulating the joint uplink scheduling and power control as a mixed discrete-continuous problem as in (20), we could replace the scheduling variable $\mathbf{s}$ with the power variable $\mathbf{p}$, based on the observation that a user $k$ is scheduled if and only if its power level $p_k$ is positive. Then, the problem can be converted to a continuous power optimization over all users. To formalize this idea, we rewrite the objective function as follows:

$$f_o(\mathbf{p}) = \sum_{i \in \mathcal{B}} \sum_{k \in \mathcal{K}_i} w_k \log \left(1 + \frac{|h_{i,k}|^2 p_k}{\sum_{k' \neq k} |h_{i,k'}|^2 p_{k'} + \sigma^2}\right) \tag{21}$$

where $k'$ refers to any other user in the network, including those who are in the same cell as user $k$, i.e., $k' \in \bigcup_{i \in \mathcal{B}} \mathcal{K}_i$. The uplink scheduling problem can then be rewritten as a power optimization problem involving only the power variable $\mathbf{p}$:

$$\underset{\mathbf{p}}{\text{maximize}} \quad f_o(\mathbf{p}) \tag{22a}$$
$$\text{subject to} \quad 0 \leq p_k \leq P_{\max}. \tag{22b}$$

Although strictly speaking, the above optimization problem does not have the constraint that at most only one user can be active, the optimal solution of (22) does take such a form in most practical regime of interest. In this case, the two problems (22) and (20) are equivalent, i.e., the optimal solution $(\mathbf{s}^\star, \mathbf{p}^\star)$ of (20) can recover the optimal solution $\mathbf{p}^\star$ of (22), and vice versa.

Problem (22) is nonconvex, but it can be solved by using the gradient method to attain a local optimum, or by using the FP method advocated in Part I of this paper [3]. After solving (22), we simply schedule those users with positive $p_k$.

However, as a subtle point we wish to highlight, using a power control algorithm to solve the scheduling problem has a serious deficiency. The main problem is that due to the highly nonconvex nature of the objective function, the stationary point of a power control algorithm is highly sensitive to the initial condition. As a result, this class of methods suffers from a serious *premature turning-off* issue. If some link is deactivated in the early stage of the iterative optimization, it can never be reactivated in the later iterations, because its local gradient would strongly discourage it from doing so. Past efforts to convexify this power control problem, e.g, by approximating the problem as a geometric program [30], essentially smooths out the local optima; but it works only at high SINR. For the scheduling problem, most of the links have low SINRs—in fact, due to intra-cell interference, at most one link in each cell can have its SINR higher than 1.

The main contribution of this paper is to show that a novel use of the Lagrangian dual transform, coupled with the quadratic transform from Part I, can avoid the premature turning-off issue through weighted bipartite matching.

*C. FP Approach*

The scheduling decision and the transmit power level of the scheduled user in each cell interact with its neighboring cells through the interference term in the denominator of rate expression in the objective function. A naive way for tackling the problem would be to make scheduling and power allocation decisions on an individual per-cell basis, assuming that the interference is fixed, then update the interference terms, and iterate between the cells. But such an approach does not work well, because the interference pattern can drastically change when a different user is scheduled; there is no guarantee that the iteration would even converge.

The main idea of this paper is to devise a way of using FP to enable the individual update of scheduling and power on a per-cell basis, while ensuring convergence. Toward this end, the quadratic transform and the Lagrangian dual transform are used together to recast the problem in a sequence of equivalent forms. We remark that applying the quadratic transform alone cannot achieve this desired decoupling.

First, apply the Lagrangian dual transform to reformulate the original objective function $f_o(\mathbf{s}, \mathbf{p})$ as

$$f_r(\mathbf{s}, \mathbf{p}, \boldsymbol{\gamma}) = \sum_{i \in \mathcal{B}} w_{s_i} \log\left(1 + \gamma_i\right) - \sum_{i \in \mathcal{B}} w_{s_i} \gamma_i + \sum_{i \in \mathcal{B}} \frac{w_{s_i}(\gamma_i + 1)|h_{i,s_i}|^2 p_{s_i}}{\sum_j |h_{i,s_j}|^2 p_{s_j} + \sigma^2} \tag{23}$$

where $\boldsymbol{\gamma}$ refers to a collection of auxiliary variables $\{\gamma_i\}_{i \in \mathcal{B}}$. The original problem (20) is now equivalent to

$$\underset{\mathbf{s}, \mathbf{p}, \boldsymbol{\gamma}}{\text{maximize}} \quad f_r(\mathbf{s}, \mathbf{p}, \boldsymbol{\gamma}) \tag{24a}$$
$$\text{subject to} \quad \text{(20b), (20c).} \tag{24b}$$

We propose to optimize all the variables iteratively. When $(\mathbf{s}, \mathbf{p})$ are held fixed, the optimal $\boldsymbol{\gamma}$ can be explicitly determined by setting $\partial f_r / \partial \gamma_i$ to zero, i.e.,

$$\gamma_i^\star = \frac{|h_{i,s_i}|^2 p_{s_i}}{\sum_{j \neq i} |h_{i,s_j}|^2 p_{s_j} + \sigma^2}. \tag{25}$$

Next, we apply the quadratic transform on the fractional term in (23) in order to to optimize $(\mathbf{s}, \mathbf{p})$ in $f_r$ for fixed $\boldsymbol{\gamma}$. Introduce an auxiliary variable $y_i$ for each ratio $\frac{w_{s_i}(\gamma_i + 1)|h_{i,s_i}|^2 p_{s_i}}{\sum_j |h_{i,s_j}|^2 p_{s_j} + \sigma^2}$ in the last term of $f_r(\mathbf{s}, \mathbf{p}, \boldsymbol{\gamma})$. We use

Corollary 1 to further reformulate $f_r(\mathbf{s}, \mathbf{p}, \boldsymbol{\gamma})$ as $f_q(\mathbf{s}, \mathbf{p}, \boldsymbol{\gamma}, \mathbf{y})$ in (26) shown at the bottom of the page. After some algebra, this $f_q$ can be rewritten in the following form:

$$f_q(\mathbf{s}, \mathbf{p}, \boldsymbol{\gamma}, \mathbf{y}) = \sum_{i \in \mathcal{B}} \left( w_{s_i} \log(1 + \gamma_i) - w_{s_i} \gamma_i - y_i^2 \sigma^2 \right.$$
$$\left. + 2y_i \sqrt{w_{s_i}(\gamma_i + 1) |h_{i,s_i}|^2 p_{s_i}} - \sum_{j \in \mathcal{B}} y_j^2 |h_{j,s_i}|^2 p_{s_i} \right) \quad (27)$$

where $\mathbf{y}$ denotes a collection of auxiliary variables $\{y_i\}_{i \in \mathcal{B}}$. Thus, in order to solve problem (24) over $(\mathbf{s}, \mathbf{p})$, we can equivalently consider the following problem over $(\mathbf{s}, \mathbf{p}, \mathbf{y})$:

$$\underset{\mathbf{s}, \mathbf{p}, \mathbf{y}}{\text{maximize}} \quad f_q(\mathbf{s}, \mathbf{p}, \boldsymbol{\gamma}, \mathbf{y}) \quad (28a)$$
$$\text{subject to} \quad (20b), (20c). \quad (28b)$$

The overall strategy is then to iteratively optimize $\boldsymbol{\gamma}$ according to (25) and optimize $(\mathbf{s}, \mathbf{p}, \mathbf{y})$ as in (28).

The newly introduced objective function $f_q$ groups the terms related to the same $s_i$ together. The key observation is that the scheduling and power variables $(\mathbf{s}, \mathbf{p})$ are now *decoupled* in this new formulation (28). Specifically, the scheduling and power optimization in each cell, i.e., $(s_i, p_i)$, can be done independently in each cell, as long as $\boldsymbol{\gamma}$ and $\mathbf{y}$ are fixed. This motivates an iterative approach for solving (28).

We propose to maximize $f_q$ over variables $\boldsymbol{\gamma}$, $\mathbf{y}$, $\mathbf{s}$ and $\mathbf{p}$ in an iterative manner as follows. The update of $\boldsymbol{\gamma}$ is already shown as in (25). When all the other variables are fixed, the optimal $\mathbf{y}$ can be obtained by setting $\partial f_q / \partial \mathbf{y}_i$ to zero, i.e.,

$$y_i^\star = \frac{\sqrt{w_{s_i}(1 + \gamma_i)|h_{i,s_i}|^2 p_{s_i}}}{\sum_{j \in \mathcal{B}} |h_{i,s_j}|^2 p_{s_j} + \sigma^2}. \quad (29)$$

Fixing $\mathbf{y}$ and $\boldsymbol{\gamma}$, if user $k$ is to be scheduled by its associated BS $j$, we can derive its optimal transmit power level $p_k$ by setting $\partial f_q / \partial p_k$ to zero. Subject to a maximum power constraints, the optimal $p_k$ can be explicitly determined by

$$p_k = \min \left\{ P_{\max}, \frac{w_k(1 + \gamma_i) |h_{i,k}|^2 y_i^2}{\left( \sum_{j \in \mathcal{B}} |h_{j,k}|^2 y_j^2 \right)^2} \right\}, \; \forall k \in \mathcal{K}_i. \quad (30)$$

The most important part of the algorithm is the optimization of the scheduling variable $\mathbf{s}$. As stated previously, the objective function $f_q$ has the desirable property that the optimization of $\mathbf{s}$ is decoupled on a per-cell basis, i.e., the optimization of $s_i$ does not depend on the other $s_j$ variables for $j \neq i$, when $\boldsymbol{\gamma}$ and $\mathbf{y}$ are fixed. Now, since the optimal transmit power level $p_k$ is already determined by (30) if user $s$ is scheduled, we can substitute the optimized power $p_k$ into $f_q$ and make optimal scheduling decision through a simple search to find the user that maximizes $f_q$ in each cell. Moreover, we can rewrite $f_q$ in the form of difference between two positive functions, and

---

**Algorithm 1: Joint Uplink Scheduling and Power Control**

**Step 0:** Initialize $\mathbf{s}$, $\mathbf{p}$ and $\boldsymbol{\gamma}$ to feasible values.
**Repeat**
    **Step 1:** Update $\mathbf{y}$ by (29).
    **Step 2:** Update $\boldsymbol{\gamma}$ by (25).
    **Step 3:** Update $(\mathbf{s}, \mathbf{p})$ jointly by (31) and (30).
**until** the value of function $f_q$ in (27) converges.

---

formally state the scheduling decision as follows:

$$s_i^\star = \begin{cases} \varnothing, \text{ if } \max_{k \in \mathcal{K}_i} \left\{ G_i(k) - \sum_{j \neq i} D_j(k) \right\} \leq 0 \\ \arg \max_{k \in \mathcal{K}_i} \left\{ G_i(k) - \sum_{j \neq i} D_j(k) \right\}, \text{ otherwise} \end{cases} \quad (31)$$

where the functions $G_i(k)$ and $D_j(k)$ are defined as

$$G_i(k) = w_k \log(1 + \gamma_i) - w_k \gamma_i - p_k y_i^2 |h_{j,k}|^2$$
$$+ 2y_i \sqrt{w_k(1 + \gamma_i) |h_{i,k}|^2 p_k}, \; \forall k \in \mathcal{K}_i \quad (32)$$

and

$$D_j(k) = y_j^2 |h_{j,k}|^2 p_k, \; \forall k \notin \mathcal{K}_j. \quad (33)$$

In the above equation (31), we interpret $G_i(k)$ and $D_j(k)$ as the utility and penalty functions, respectively, so that the scheduling decision has an intuitive utility-minus-price structure. More precisely, $G_i(k)$ is the utility gain of scheduling user $k$ at BS $i$ and $D_j(k)$ is the penalty for interfering a neighboring cell $j$ by scheduling user $k$. The best user to schedule is the one that balances these two effects. Note that the scheduling and power control are done on a per-cell basis. This enables distributed implementation.

Furthermore, when the max value of $G_i(k) - \sum_{j \neq i} D_j(k)$ at BS $i$ is less than zero, it implies that no user should be scheduled at this BS $i$ in the time slot in order to reduce the intercell interference suffered by the neighboring BSs. This case possibly happens in an ultra-dense uplink network scenario.

We summarize the proposed joint scheduling and power control strategy in Algorithm 1. Note that the algorithm is not a conventional block coordinate ascent method, because the optimizing objective function is not fixed, i.e., $\mathbf{s}$, $\mathbf{p}$ and $\mathbf{y}$ are optimally updated for $f_q$ while $\boldsymbol{\gamma}$ is optimally updated for $f_r$. Nevertheless, its convergence can be established, as specified in Proposition 1.

*Proposition 1:* Algorithm 1 is guaranteed to converge, with the weighted sum rate $f_o$ monotonically nondecreasing after each iteration. The converged solution is a stationary point of

---

$$f_q(\mathbf{s}, \mathbf{p}, \boldsymbol{\gamma}, \mathbf{y}) = \sum_{i \in \mathcal{B}} w_{s_i} \log(1 + \gamma_i) - \sum_{i \in \mathcal{B}} w_{s_i} \gamma_i + \sum_{i \in \mathcal{B}} \left( 2y_i \sqrt{w_{s_i}(\gamma_i + 1) |h_{i,s_i}|^2 p_{s_i}} - y_i^2 \left( \sum_{j \in \mathcal{B}} |h_{i,s_j}|^2 p_{s_j} + \sigma^2 \right) \right) \quad (26)$$

TABLE I
SUM LOG-UTILITIES OF FP-BASED COORDINATED UPLINK SCHEDULING
AND POWER CONTROL AS COMPARED TO THE BASELINES

| Algorithm | Total log-utility |
|---|---|
| Power Control by WMMSE | 27.17 |
| Fixed Interference | 52.16 |
| Proposed FP Method | 60.15 |

$f_o$ with respect to **p** if **s** is assumed to be fixed.

*Proof:* See Appendix A. ∎

We note that due to the nonconvex nature of the problem with respect to **p**, finding a stationary point in **p** is likely to be the best that one can do. Moreover, since **s** is a discrete variable, it is difficult to assert any optimality with respect to **s**. In fact, we can show that even with **p** fixed, finding the optimal **s** is NP-hard.

To see the NP-hardness, we can use an argument inspired by [31] in which the NP-hardness of the power control problem is established. Construct a simplified example, in which each BS receives interference from a subset of neighboring users only, and the interference level is large so whenever interference is present the rate is effectively zero, and otherwise the rate is one. Selecting one user in each cell to maximize the overall sum rate now amounts to solving a maximum independent set problem on a graph, which is NP-hard. Further, unless P = NP, it is impossible even to solve the problem within a constant approximation ratio in polynomial time [32].

Observe here that Algorithm 1 avoids premature turning-off. Even if a user $k$ is not activated in the $t$th iterate, the related auxiliary variable $y_i$ is still non-zero according to (29), so long as at least some other user is scheduled in its cell. Thus, user $k$ still stands a chance to be reactivated in future iterations when the interference pattern becomes favorable, as indicated by (30).

As a final remark, throughout this paper we have assumed the availability of CSI for uplink scheduling. In practical implementations, the cost of obtaining CSI for all users can be prohibitive. Further, including all users in the scheduling step can incur large computational complexity. The complexity in implementing Algorithm 1 can be lowered in practice using a two-stage scheduling strategy. We first roughly choose a subset of potential users according to their weights, then apply Algorithm 1 to refine the scheduling decision. This can greatly reduce the run-time complexity and the cost of obtaining CSI.

*D. Simulation Results*

To evaluate the performance of the proposed joint uplink scheduling and power control algorithm, numerical simulation is performed in a 7-cell wrapped-around topology with a total of 84 users uniformly placed in the network. The BS-to-BS distance is 800m. Each user is associated with the strongest BS. The maximum transmit power spectrum density (PSD) of the users is $-47$dBm/Hz; the background noise PSD is set to be $-169$dBm/Hz over 10MHz bandwidth. The wireless channel model includes a distance-dependent pathloss component at $128.1 + 37.6 \log_{10}(d)$dB (where the distance $d$

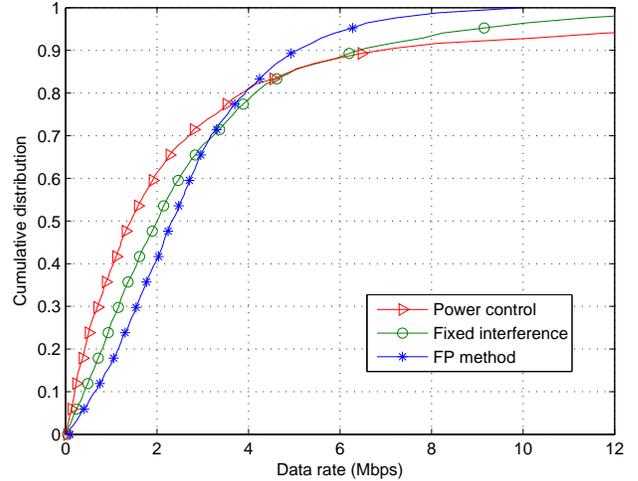

Fig. 2. Comparison of the proposed FP-based coordinated uplink user scheduling and power control method with two baseline methods in term of cumulative distribution function of user rates.

is in km) and a log-normal shadowing component with 8dB standard deviation.

In the simulation, the joint user scheduling and power control problem is solved across the multiple cells in each timeslot with the user priority weights updated as the reciprocals of long-term average user rates over the time, in order to ensure proportional fairness across the users. Over time, this setting of the weights maximizes the log-utility, $\sum_k \log(\bar{R}_k)$, over all users in the network, where $\bar{R}_k$ is the long-term average rate of user $k$, expressed in Mbps in the numerical results below.

The following two baseline uplink scheduling strategies are also simulated for comparison purpose:

- *Power Control*: The uplink scheduling and power control problem can also be thought of as a global power control problem, in which users not being scheduled are assigned zero power. Thus, we can run power control for all the users in the network at the same time. Most users will be assigned zero power; users assigned positive transmit power levels (typically at most one per cell) are the ones scheduled. This global power control problem is highly nonconvex. In the simulation, we use the WMMSE algorithm [27], [28] for power control to arrive at a local optimum.
- *Fixed Interference Method*: In this method, uplink scheduling and power control are performed iteratively. Each user is initialized with some power level. In the scheduling stage, the user that maximizes the weighted rate in each cell is chosen, assuming fixed interference pattern from the previous iteration. In the power control stage, the powers of the scheduled users are updated by solving a weighted sum rate maximization problem. We iterate between the two steps until convergence or a fixed number of iterations is reached.

Fig. 2 shows the cumulative distribution of the user data rates in the network and Table I lists the log-utility[2] achieved by the different methods for uplink user scheduling and power control. We see that the baseline of power control provides poor performance, mainly because the power control algorithm tends to stuck in a locally optimal solution of the nonconvex problem. The fixed-interference method is also not capable of arriving at a desirable solution. In contrast, the proposed algorithm performs much better in terms of utility, as shown in Table I. Fig. 2 shows that the 10th-percentile user rate of the proposed algorithm is at least 50% more than that of the fixed-interference method. Since these low-rate users are typically located at the cell-edge where cross-cell interference is the strongest, this shows that the proposed FP-based algorithm is effective in alleviating interference by coordinating uplink scheduling and power control. We remark that this gain is achieved despite the low overall complexity of the FP method. A detailed complexity analysis is included in the next section.

---
[2]The utility is computed for data rate in Mbps throughout the paper.

## V. Joint Uplink Scheduling and Beamforming

We now consider a more general problem for the uplink of a MIMO multicell network, where the transmit beamformers are optimized in addition to user schedule and power.

### A. Problem Formulation

Following the notations as in Section IV, define $\mathcal{B}$ as the set of BSs in the network, $\mathcal{K}_i$ as the set of users who are associated with BS $i$, $\sigma^2$ as the background noise level, $w_k$ as the weight of user $k$, and $P_{\max}$ as the maximum transmit power level at the user side. Assume that each user is equipped with $N$ antennas and each BS is equipped with $M$ antennas. Spatial multiplexing can therefore support up to $M$ data streams per cell (but some data streams may have zero throughput). Let $s_{im}$ be the index of the user who is scheduled in the $m$th stream at BS $i$. Let $\mathbf{v}_k \in \mathbb{C}^N$ be the transmit beamformer of user $k$ if it gets scheduled. Let $\mathbf{H}_{i,k} \in \mathbb{C}^{M \times N}$ be the uplink channel from user $k$ to BS $i$. The joint uplink user scheduling and beamforming problem with a weighted sum-rate maximizing objective can be formulated as

$$\underset{\mathbf{s}, \mathbf{V}}{\text{maximize}} \quad f_o(\mathbf{s}, \mathbf{V}) \tag{34a}$$
$$\text{subject to} \quad \|\mathbf{v}_{im}\|_2^2 \leq P_{\max} \tag{34b}$$
$$s_{im} \in \mathcal{K}_i \cup \{\varnothing\} \tag{34c}$$

with the objective function $f_o$ defined in (35) at the bottom of the page, where $\mathbf{s}$ denotes $\{s_{im}\}_{i \in \mathcal{B}, m \in \{1,...,M\}}$ and $\mathbf{V}$ denotes $\{\mathbf{V}_{im}\}_{i \in \mathcal{B}, m \in \{1,...,M\}}$. Note that under this MIMO setting we allow scheduling up to $M$ users per cell.

The above problem is more challenging than the uplink user scheduling and power control problem (20) of the SISO case. In addition to the crosscell interference, we also need to take into account the interference coming from the same cell because multiple users can be scheduled at each BS.

### B. FP Reformulation and Weighted Bipartite Matching

Recall that in Section IV-C we make use of the quadratic transform and the Lagrangian dual transform to derive a reformulation for the joint uplink scheduling and power control problem, whereby the power and scheduling variables can be grouped on a per-cell basis. This reformulating procedure can be adapted to the multidimensional case for problem (34). First, apply the multidimensional Lagrangian dual transform in Theorem 4 to reformulate the original objective function $f_o(\mathbf{s}, \mathbf{V})$ as $f_r(\mathbf{s}, \mathbf{V}, \boldsymbol{\gamma})$ in (36) shown at the bottom of this page, with an auxiliary variable $\gamma_{im}$ introduced for each data stream $(i, m)$, and the collection $\{\gamma_{im}\}$ denoted by $\boldsymbol{\gamma}$. Thus, the original problem (34) is equivalent to

$$\underset{\mathbf{s}, \mathbf{V}, \boldsymbol{\gamma}}{\text{maximize}} \quad f_r(\mathbf{s}, \mathbf{V}, \boldsymbol{\gamma}) \tag{37a}$$
$$\text{subject to} \quad (34b), (34c). \tag{37b}$$

Following Algorithm 1, we propose to optimize the variables in (37) in an iterative fashion. When the primal variables $\mathbf{s}$ and $\mathbf{v}$ are both held fixed, maximizing $f_r$ over $\boldsymbol{\gamma}$ is a convex problem which can be efficiently solved by setting $\partial f_r / \partial \gamma_{im}$ to zero, that is shown in (38) at the bottom of the page. Note that the optimal $\gamma_{im}$ is equal to the resulting uplink SINR in data stream $(i, m)$ exactly.

We then consider optimizing $\mathbf{s}$ and $\mathbf{V}$ for fixed $\boldsymbol{\gamma}$. This subproblem only involves the last term of $f_q$ which has a multidimensional sum-of-ratio form. By treating $\sqrt{w_{s_{im}}(1+\gamma_{im})}\mathbf{H}_{i,s_{im}}\mathbf{v}_{s_{im}}$ as the numerator vector $\boldsymbol{\alpha}$ and $\left(\sigma^2 \mathbf{I} + \sum_{(j,n)} \mathbf{H}_{i,s_{jn}} \mathbf{v}_{s_{jn}} \mathbf{v}_{s_{jn}}^\dagger \mathbf{H}_{i,s_{jn}}^\dagger\right)$ as the denominator

$$f_o(\mathbf{s}, \mathbf{V}) = \sum_{(i,m)} w_{s_{im}} \log\left(1 + \mathbf{v}_{s_{im}}^\dagger \mathbf{H}_{i,s_{im}}^\dagger \left(\sigma^2 \mathbf{I} + \sum_{(j,n) \neq (i,m)} \mathbf{H}_{i,s_{jn}} \mathbf{v}_{s_{jn}} \mathbf{v}_{s_{jn}}^\dagger \mathbf{H}_{i,s_{jn}}^\dagger\right)^{-1} \mathbf{H}_{i,s_{im}} \mathbf{v}_{s_{im}}\right) \tag{35}$$

$$f_r(\mathbf{s}, \mathbf{V}, \boldsymbol{\gamma}) = \sum_{(i,m)} w_{s_{im}} \left(\log(1+\gamma_{im}) - \gamma_{im} + (1+\gamma_{im})\mathbf{v}_{s_{im}}^\dagger \mathbf{H}_{i,s_{im}}^\dagger \left(\sigma^2 \mathbf{I} + \sum_{(j,n)} \mathbf{H}_{i,s_{jn}} \mathbf{v}_{s_{jn}} \mathbf{v}_{s_{jn}}^\dagger \mathbf{H}_{i,s_{jn}}^\dagger\right)^{-1} \mathbf{H}_{i,s_{im}} \mathbf{v}_{s_{im}}\right) \tag{36}$$

$$\gamma_{im}^\star = \mathbf{v}_{s_{im}}^\dagger \mathbf{H}_{i,s_{im}}^\dagger \left(\sigma^2 \mathbf{I} + \sum_{(j,n) \neq (i,m)} \mathbf{H}_{i,s_{jn}} \mathbf{v}_{s_{jn}} \mathbf{v}_{s_{jn}}^\dagger \mathbf{H}_{i,s_{jn}}^\dagger\right)^{-1} \mathbf{H}_{i,s_{im}} \mathbf{v}_{s_{im}} \tag{38}$$

matrix $\mathbf{B}$ in Theorem 2, we arrive at a new objective

$$f_q(\mathbf{s}, \mathbf{V}, \boldsymbol{\gamma}, \mathbf{Y}) = \sum_{(i,m)} w_{s_{im}} \log(1+\gamma_{im}) - \sum_{(i,m)} w_{s_{im}} \gamma_{im}$$
$$+ \sum_{(i,m)} \left( 2\sqrt{w_{s_{im}}(1+\gamma_{im})} \operatorname{Re}\left\{ \mathbf{v}_{s_{im}}^\dagger \mathbf{H}_{i,s_{im}}^\dagger \mathbf{y}_{im} \right\} \right.$$
$$\left. - \mathbf{y}_{im}^\dagger \left( \sigma^2 \mathbf{I} + \sum_{(j,n)} \mathbf{H}_{i,s_{jn}} \mathbf{v}_{s_{jn}} \mathbf{v}_{s_{jn}}^\dagger \mathbf{H}_{i,s_{jn}}^\dagger \right) \mathbf{y}_{im} \right) \quad (39)$$

where an auxiliary variable $\mathbf{y}_{im} \in \mathbb{C}^M$ is introduced with respect to each data stream $(i,m)$, and the collection of auxiliary variable $\{\mathbf{y}_{im}\}$ is denoted by $\mathbf{Y}$. Thus, the optimization of $f_r$ in (37) is further recast to

$$\underset{\mathbf{s}, \mathbf{V}, \boldsymbol{\gamma}, \mathbf{Y}}{\text{maximize}} \quad f_q(\mathbf{s}, \mathbf{V}, \boldsymbol{\gamma}, \mathbf{Y}) \quad (40a)$$
$$\text{subject to} \quad (34b), (34c). \quad (40b)$$

With the update of $\boldsymbol{\gamma}$ already shown in (38), we now consider the optimization of $\mathbf{s}$, $\mathbf{V}$ and $\mathbf{Y}$ in $f_q$. First, when all the other variables are fixed, the optimal $\mathbf{y}$ can be explicitly determined by setting $\partial f_r / \partial \mathbf{y}_{im}$ to zero, that is

$$\mathbf{y}_{im}^\star = \left( \sigma^2 \mathbf{I} + \sum_{(j,n)} \mathbf{H}_{i,s_{jn}} \mathbf{v}_{s_{jn}} \mathbf{v}_{s_{jn}}^\dagger \mathbf{H}_{i,s_{jn}}^\dagger \right)^{-1}$$
$$\cdot \sqrt{w_{s_{im}}(1+\gamma_{im})} \mathbf{H}_{i,s_{im}} \mathbf{v}_{s_{im}}. \quad (41)$$

Observe that the optimal $\mathbf{y}_{im}$ is exactly a minimum mean-square-error (MMSE) receiver scaled by a factor of $\sqrt{w_{s_{im}}(1+\gamma_{im})}$, with respect to each data stream $(i,m)$.

It remains to optimize the variables $\mathbf{s}$ and $\mathbf{V}$ in $f_q$. We gain incorporate the idea of *weighted bipartite matching* for the joint optimization of these two variables. The key observation is that the scheduling of user $s_{im}$ and its transmit beamformer $\mathbf{v}_{im}$ in a particular data stream $(i,m)$ contribute to the objective function (39) in a way that is *independent* of the scheduling and beamformer choices in other streams. More specifically, if some user $k$ is scheduled in the data stream $(i,m)$, i.e., $s_{im} = k$, then the optimal transmit beamformer of user $k$ with respect to $(i,m)$, denoted as $\boldsymbol{\tau}_{k,im}$, can be determined by solving $\partial f_q / \partial \mathbf{v}_{s_{im}} = \mathbf{0}$, i.e.,

$$\boldsymbol{\tau}_{k,im} = \left( \sum_{(j,n)} \mathbf{H}_{j,k}^\dagger \mathbf{y}_{jn} \mathbf{y}_{jn}^\dagger \mathbf{H}_{j,k} + \eta_{k,im}^\star \mathbf{I} \right)^{-1}$$
$$\cdot \sqrt{w_k(1+\gamma_{im})} \mathbf{H}_{i,k}^\dagger \mathbf{y}_{im}. \quad (42)$$

where the dual variable $\eta_{k,im}^\star$ accounts for power constraint (34b) and is optimally determined by the complementary slackness condition

$$\eta_{k,im}^\star = \min\{\eta_{k,im} \geq 0 : \|\boldsymbol{\tau}_{k,im}(\eta_{k,im})\|_2^2 \leq P_{\max}\}. \quad (43)$$

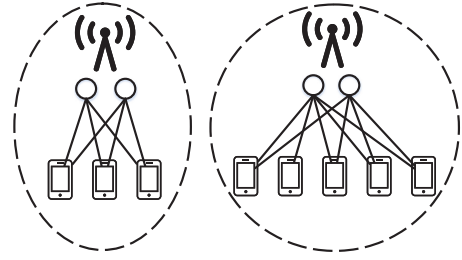

Fig. 3. The scheduling variable $\mathbf{s}$ is decoupled on a per-cell basis after the FP-based reformulation. Optimizing the scheduling variable $\mathbf{s}$ in (40) can be characterized as a weighted bipartite matching between the users and the data streams in each cell, with the matching weights defined by (44).

This $\eta_{k,im}^\star$ can be efficiently evaluated via bisection search.

Therefore, the utility value (in terms of $f_q$) of scheduling user $k$ in one particular data stream $(i,m)$ can be determined analytically. This allows solving $\mathbf{s}$ and $\mathbf{V}$ jointly by weighted bipartite matching. To formalize the idea, we define the utility value of assigning user $k$ to data stream $(i,m)$ as $\xi_{k,im}$ in (44) at the bottom of the page.

Then, the $f_q$ maximizing problem (40) reduces to the following weighted bipartite matching problem:

$$\underset{\mathbf{x}}{\text{maximize}} \quad \sum_{k \in \mathcal{K}_i} \sum_{m=1}^{N} \xi_{k,im} x_{k,im} \quad (45a)$$
$$\text{subject to} \quad \sum_{k \in \mathcal{K}_i} x_{k,im} \leq 1, \forall m \quad (45b)$$
$$\sum_{m=1}^{N} x_{k,im} \leq 1, \forall k \quad (45c)$$
$$x_{k,im} \in \{0, 1\}, \quad (45d)$$

where the binary variable $x_{k,im}$ indicates whether or not user $k$ is scheduled in the $m$th data stream at its associated BS $i$. We remark that the above matching problem is considered at each BS $i$ *individually*, as illustrated in Fig. 3.

Weighted bipartite matching is a well-studied problem in the field of combinatorics [33]. It can be efficiently solved by the existing algorithms with polynomial-time computational complexity using, e.g., the Hungarian algorithm [6] and the auction algorithm [7], with a computational complexity of $O((K+M)^3)$. Further, because in practice the matching weights $\xi_{k,im}$ are always evaluated with finite precision, in this finite-precise case, the complexity of matching can be reduced to $O((K+M)^2)$ using the algorithm in [34].

After solving for $\mathbf{x}$ in problem (45), we recover the optimal scheduling variable $\mathbf{s}^\star$ by

$$s_{im}^\star = \begin{cases} k, & \text{if } x_{k,im}^\star = 1 \text{ for some } k \in \mathcal{K}_i \\ \varnothing, & \text{otherwise} \end{cases} \quad (46)$$

where the decision $\varnothing$ is made in data stream $(i,m)$ if any

$$\xi_{k,im} = w_k \log(1+\gamma_{im}) - w_k \gamma_{im} + 2\sqrt{w_k(1+\gamma_{im})} \operatorname{Re}\left\{ \boldsymbol{\tau}_{k,im}^\dagger \mathbf{H}_{i,k}^\dagger \mathbf{y}_{im} \right\} - \sigma^2 \|\mathbf{y}_{im}\|_2^2 - \sum_{(j,n)} \mathbf{y}_{jn}^\dagger \mathbf{H}_{j,k} \boldsymbol{\tau}_{k,im} \boldsymbol{\tau}_{k,im}^\dagger \mathbf{H}_{j,k}^\dagger \mathbf{y}_{jn} \quad (44)$$

| **Algorithm 2: Joint Uplink Scheduling and Beamforming** |
|---|
| **Step 0:** Initialize $\mathbf{s}$, $\mathbf{V}$ and $\boldsymbol{\gamma}$ to feasible values. |
| **Repeat** |
|     **Step 1:** Update $\mathbf{Y}$ by (41). |
|     **Step 2:** Update $\boldsymbol{\gamma}$ by (38). |
|     **Step 3:** Update $(\mathbf{s}, \mathbf{V})$ jointly by (42), (46) and (47). |
| **until** the value of function $f_q$ in (39) converges. |

user scheduled in the stream would have contributed to $f_q$ negatively. Note that $x_{k,im}^\star$ must be zero if $\xi_{k,im} < 0$. In practice, we can further facilitate weighted matching by removing the edges corresponding to negative $\xi_{k,im}$ from the bipartite graph. The transmit beamformers of the scheduled users are then set to the optimal values in (42) accordingly:

$$\mathbf{v}_k^\star = \boldsymbol{\tau}_{k,im}, \quad \text{if } x_{k,im}^\star = 1 \text{ for some } (i,m). \qquad (47)$$

We summarize the proposed iterative distributed optimization in Algorithm 2.

Like Algorithm 1, this algorithm guarantees convergence although it is not a block coordinate ascent method, as stated in the following proposition.

*Proposition 2:* Algorithm 2 is guaranteed to converge, with the weighted sum rate $f_o$ monotonically nondecreasing after each iteration. The converged solution is a stationary point of $f_o$ with respect to $\mathbf{V}$ if $\mathbf{s}$ is assumed to be fixed.

    *Proof:* See Appendix A. ∎

We note that the SISO algorithm in Section IV is a special case of the weighted bipartite matching approach for the MIMO problem. Further, we can use the same agument to show that computing the optimal $\mathbf{s}$ for fixed $\mathbf{V}$ is already NP-hard, so the above convergence result is likely the best one can hope for.

As a final remark, Algorithms 1 and 2 can be initialized with simple but reasonable heuristic. For example, in a $2 \times 2$ MIMO network, the two users with the highest weights in each cell can be scheduled at the beginning, and their beamformers can be set to maximize the signal strength. Moreover, we set some small constant $\delta > 0$ and use the convergence criterion $|f_q^{(t)} - f_q^{(t-1)}| < \delta$ where $t$ is the iteration index.

### C. Discrete Beamforming

So far it is assumed that each beamformer $\mathbf{v}_{im}$ can be set to an arbitrary vector as long as the power constraint $\|\mathbf{v}_{im}\|_2^2 \leq P_{\max}$ is satisfied. We now consider a discrete scenario for beamforming where the choice for $\mathbf{v}_{im}$ is restricted to a codebook

$$\mathcal{V} = \{\boldsymbol{\phi}_1, \boldsymbol{\phi}_2, \cdots, \boldsymbol{\phi}_{|\mathcal{V}|}\}. \qquad (48)$$

In the above, each $\boldsymbol{\phi}_n \in \mathbb{C}^N$ (for $n = 1, \ldots, |\mathcal{V}|$) represents a possible beamforming vector.

In this case, if some user $k$ is scheduled in the data stream $(i, m)$, then its optimal transmit beamformer $\boldsymbol{\tau}_{k,im}$ in terms of $f_q$ can be obtained by searching through the codebook, i.e.,

$$\boldsymbol{\tau}_{k,im} = \arg\max_{\mathbf{v} \in \mathcal{V}} \left\{ 2\sqrt{w_k(1+\gamma_{im})} \operatorname{Re}\left\{\mathbf{v}^\dagger \mathbf{H}_{i,k}^\dagger \mathbf{y}_{im}\right\} \right.$$
$$\left. - \sum_{(j,n)} \mathbf{y}_{jn}^\dagger \mathbf{H}_{j,k} \mathbf{v} \mathbf{v}^\dagger \mathbf{H}_{j,k}^\dagger \mathbf{y}_{jn} \right\}. \qquad (49)$$

The bipartite matching process (45) can then be performed with $\xi_{k,im}$ set according to (44) but using the above $\boldsymbol{\tau}_{k,im}$. After matching, the optimal $\mathbf{V}$ is recovered by (47).

To find the optimal $\mathbf{v}_{im}$ in the discrete search (49) requires at most a computational complexity of $O(|\mathcal{V}|)$. This complexity can be further reduced to $O(\log |\mathcal{V}|)$ by taking advantage of the functional structure of (49). The idea is to first maximize $f_q$ over $\mathbf{V}$ without considering the discrete constraint (48) and then find the discrete solution $\boldsymbol{\phi} \in \mathcal{V}$ that is closest to the relaxed solution $\tilde{\mathbf{v}}_{im}$, for every $(i, m)$ pair, i.e.,

$$\boldsymbol{\tau}_{k,im} = \arg\min_{\boldsymbol{\phi} \in \mathcal{V}} \|\boldsymbol{\phi} - \tilde{\mathbf{v}}_{im}\|_2 \qquad (50)$$

where the relaxed solution $\tilde{\mathbf{v}}_{im}$ is the $\boldsymbol{\tau}_{k,im}$ in (42) without the discrete codebook constraint. Observe that the right-hand side of (49) is a concave quadratic function of variable $\mathbf{v}$, and then after completing the square, it can be shown that updating $\boldsymbol{\tau}_{k,im}$ by (50) yields exactly the same solution as in (49). Therefore, although the above relax-and-then-round approach in (50) is a common heuristic for discrete beamforming, our FP framework gives a theoretical justification by showing that this approach actually maximizes the reformulated objective $f_q$, which acts as a lower bound of the original objective $f_o$ according to Lemmas 1 and 2 in Appendix A.

An efficient way to perform the optimization (49) can now be devised based on (50), as stated in the following proposition.

*Proposition 3 (Nearest Point Projection for Beamforming):* The optimal update (49) for discrete beamforming can be realized by the nearest point projection as in (50) with a computational complexity of $O(\log |\mathcal{V}|)$.

    *Proof:* Construct a $k$-d tree [35] for all the elements of $\mathcal{V}$ in advance. The following three steps produce the nearest-point projection (50): Insert $\tilde{\mathbf{v}}_{im}$ in the $k$-d tree; then search for the nearest neighbor of $\tilde{\mathbf{v}}_{im}$ in the tree and output it as the projection result; finally delete $\tilde{\mathbf{v}}_{im}$ from the tree. The insertion, search, and deletion operations all have an average complexity $O(\log |\mathcal{V}|)$. ∎

We remark that a similar result can be derived for discrete power control in the SISO case, in which case the search through the $k$-d tree reduces to a one-dimensional bisection search.

### D. Simulation Results

We validate the proposed FP-based approach by simulating a network consisting of 7 cells in a wrapped around topology. A total of 84 users randomly distributed in the network are associated with the BS to which the channel is the strongest. Each user is equipped with 2 antennas and each BS is equipped with 4 antennas. The uplink MIMO channels consist of two components: a large-scale fading component (i.e., pathloss and shadowing), which follows the model discussed in Section IV-D, and a Rayleigh fading component. The user weights in every time-slot are updated as the reciprocal of the long-term average rates in order to maximize a proportional fairness utility. All other parameters, i.e., the channel pathloss model, background noise level, maximum transmit power level, and bandwidth, follow the settings in Section IV-D.

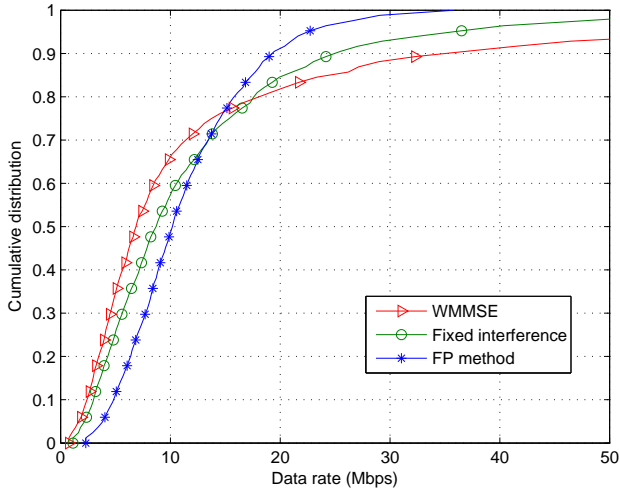

Fig. 4. Comparison of the proposed FP-based coordinated uplink user scheduling and beamforming method with two baseline methods in terms of cumulative distribution function of user rates.

TABLE II
SUM LOG-UTILITIES OF THE PROPOSED COORDINATED UPLINK SCHEDULING AND BEAMFORMING METHOD AS COMPARED TO THE TWO BASELINE SCHEMES

| Algorithm | Total log-utility |
| --- | --- |
| WMMSE | 175.87 |
| Fixed Interference | 183.45 |
| Proposed FP Method | 193.79 |

The following methods are introduced as benchmarks:

- *WMMSE*: The WMMSE algorithm is introduced in [27], [28]. To use WMMSE for user scheduling, we initialize all the users in the network with some random beamformers, then run the WMMSE algorithm to optimize weighted sum rate. At convergence, most users would be assigned zero beamformer; those assigned nonzero beamformers are scheduled. User scheduling is therefore determined *implicitly* by beamforming. In the SISO case, the beamforming step reduces to power control.
- *Fixed interference method*: This heuristic method extends the fixed interference method in Section IV-D. Iteratively, apply a beamforming method (e.g., WMMSE) for fixed user scheduling variable $\mathbf{s}$, and then optimize $\mathbf{s}$ for fixed beamformers. This works well in the downlink because the optimal scheduling can be explicitly determined [12]. For the uplink, the heuristic is to emulate the downlink by assuming fixed interference from the neighboring cells.

The proposed algorithm is compared with the aforementioned two baselines. As shown in Fig. 4, the proposed FP-based method has a significant advantage over the baselines particularly for low-rate users. For example, the rates of the 10th-percentile users is improved by at least 50% the proposed algorithm. These low-rate users are mostly located close to the cell edges, highlighting the important role of coordinated uplink scheduling and beamforming in interference mitigation. Table II shows that the proposed FP method substantially improves the sum log-utility in the network as compared to the benchmarks, verifying that interference management by coordinating user schedules and beamformers is crucial to the network performance.

## VI. CONNECTION WITH WMMSE

As already mentioned, the well-known WMMSE algorithm [27], [28] can already be used for the uplink coordinated joint scheduling, power control, and beamforming problem. Assume that all the users in the network are scheduled at the beginning; run the WMMSE algorithm to design beamformers for all the users; then only schedule the users with positive transmit power levels at the end. Interestingly, there is in fact a connection between WMMSE and our FP approach.

### A. Interpretation of WMMSE from FP

The WMMSE algorithm is originally derived based on a signal minimum mean-square-error analysis [27], [28]. In what follows, we give another derivation for WMMSE based on the proposed quadratic transform. Recall that after the use of Lagrangian dual transform, the original objective function $f_o(\mathbf{s}, \mathbf{V})$ is recast to $f_r(\mathbf{s}, \mathbf{V}, \boldsymbol{\gamma})$, in which the primal variables $\mathbf{s}$ and $\mathbf{V}$ only appear in the last sum-of-ratio term. Specifically, each ratio contained in the sum-of-ratio term of $f_r$ can be written as

$$d_{im}\mathbf{v}_{s_{im}}^\dagger \mathbf{H}_{i,s_{im}}^\dagger \mathbf{B}_{im}^{-1} \mathbf{H}_{i,s_{im}} \mathbf{v}_{s_{im}} \tag{51}$$

where two new notations $d_{im}$ and $\mathbf{B}_{im}$ are introduced to simplify notation:

$$d_{im} = w_{s_{im}}(1 + \gamma_{s_{im}}) \tag{52}$$

and

$$\mathbf{B}_{im} = \sigma^2 \mathbf{I} + \sum_{(j,n)} \mathbf{H}_{i,s_{jn}} \mathbf{v}_{s_{jn}} \mathbf{v}_{s_{jn}}^\dagger \mathbf{H}_{i,s_{jn}}^\dagger. \tag{53}$$

Recall that in deriving the further reformulation of $f_q$, we propose in Section V-B to apply the multidimensional quadratic transform in Theorem 2 by identifying the ratio pattern of (51) as

$$\left(\sqrt{d_{im}}\mathbf{H}_{i,s_{im}}\mathbf{v}_{s_{im}}\right)^\dagger \mathbf{B}_{im}^{-1} \underbrace{\left(\sqrt{d_{im}}\mathbf{H}_{i,s_{im}}\mathbf{v}_{s_{im}}\right)}_{\text{Numerator vector }\boldsymbol{\alpha}} \tag{54}$$

where $\boldsymbol{\alpha}$ represents the numerator vector in the multidimensional FP problem.

However, this is not the only way to implement the FP technique. In fact, we could have applied the multidimensional quadratic transform to the ratios in a different way:

$$d_{im}\left((\mathbf{H}_{i,s_{im}}\mathbf{v}_{s_{im}})^\dagger \mathbf{B}_{im}^{-1} \underbrace{(\mathbf{H}_{i,s_{im}}\mathbf{v}_{s_{im}})}_{\text{Numerator vector }\boldsymbol{\alpha}}\right). \tag{55}$$

In this case, we would have arrived at a different reformulation $\check{f}_q$ as shown in (56) at the bottom of the page.

This reformulation gives the following iterative algorithm for optimizing beamformers. Finding the optimal $\mathbf{Y}$ by solving $\partial \check{f}_q/\partial \mathbf{y}_{im} = \mathbf{0}$ with respect to each $(i,m)$ pair amounts to

$$\check{\mathbf{y}}_{im} = \left(\sigma^2 \mathbf{I} + \sum_{(j,n)} \mathbf{H}_{i,s_{jn}} \mathbf{v}_{s_{jn}} \mathbf{v}_{s_{jn}}^\dagger \mathbf{H}_{i,s_{jn}}^\dagger \right)^{-1} \mathbf{H}_{i,s_{im}} \mathbf{v}_{s_{im}}. \tag{57}$$

Note that the above $\check{\mathbf{y}}_{im}$ solution is exactly an MMSE receiver. Likewise, the optimal transmit beamformer is

$$\check{\mathbf{v}}_{s_{im}} = \left( \sum_{(j,n)} d_{jn} \mathbf{H}_{j,s_{im}}^\dagger \mathbf{y}_{jn} \mathbf{y}_{jn}^\dagger \mathbf{H}_{j,s_{im}} + \eta_{im}^\star \mathbf{I} \right)^{-1} \cdot d_{im} \mathbf{H}_{i,s_{im}}^\dagger \mathbf{y}_{im} \tag{58}$$

where $\eta_{im}^\star = \min\{\eta_{im} \geq 0 : \|\check{\mathbf{v}}(\eta_{im})\|_2^2 \leq P_{\max}\}$ is the optimal dual variable for the power constraint (34b) by complementary slackness. Finally, the update of $\boldsymbol{\gamma}$ remains the same as in (38). When iteratively applying the above updates of $\boldsymbol{\gamma}$, $\mathbf{V}$ and $\mathbf{Y}$ for the fixed scheduling variable $\mathbf{s}$, we arrive at exactly the WMMSE algorithm for beamforming. Therefore, WMMSE can be interpreted as a specific way of using FP to solve the optimal beamforming problem.

However, unlike our proposed reformulation $f_q$ in (39), this $\check{f}_q$ does not allow an explicit distributed solution for $\mathbf{s}$, because the discrete variables $s_i$'s are *not decoupled* in the last term of $\check{f}_q$ as shown in (56). While the FP-based method proposed in this paper is able to use weighted bipartite matching to find the optimal $\mathbf{s}$, the WMMSE algorithm can only optimize the scheduling variable implicitly by optimizing beamformers for all the users in the network. This implicit scheduling of WMMSE is not only more computationally complex, but also has inferior performance as shown in the previous section.

### B. Complexity Comparison

We now compare the complexities of Algorithm 2 and the WMMSE method [27], [28] (which is modified to include scheduling as stated in Section VI). For ease of analysis, assume that each cell has the same number of users. Let $K$ be the number of users per cell; let $B$ be the total number of BSs deployed throughout the network. Following [28], we evaluate the algorithm complexity with respect to each round of iteration.

First consider the *communication complexity*. In Algorithm 2, every BS needs to collect $(\mathbf{s}, \mathbf{V}, \mathbf{Y})$ except $\boldsymbol{\gamma}$ with respect to each $(j,n)$ pair, so the overall communication complexity of Algorithm 2 is $O(M^2 B^2 + MNB^2)$, which is independent of $K$. In the WMMSE method, each BS needs to collect $(\mathbf{V}, \boldsymbol{\gamma}, \mathbf{Y})$ with respect to every user in the network, thus the overall communication complexity of WMMSE is $O(MKB^2 + NKB^2)$. WMMSE in general has a much higher communication complexity, because normally $K \gg M$ (i.e., only a small portion of users in the cell are scheduled in each time-slot).

We further analyze the *computational complexity*. Assuming that the classic Hungarian algorithm is used for weighted bipartite matching, the overall computational complexity of Algorithm 2 can be shown to be $O(c_{\text{FP}})$, where $c_{\text{FP}} = M^4 B^2 + MN^3 KB + (M^3 N + MN^2) KB^2 + (K + M)^3 B$. The WMMSE algorithm involves a matrix multiplication with respect to every user-BS pair in the network. Consequently, it requires a computational complexity of $O(c_{\text{WMMSE}})$, where $c_{\text{WMMSE}} = (M^3 + N^3) KB + M^2 KB^2 + (MN + N^2) K^2 B^2$. We remark that *matrix chain ordering* needs to be optimized for both of the algorithms to find the most efficient way of multiplying matrices. For simplicity, we further assume that $M$ and $N$ are fixed and also that $K$ is much greater than both $M$ and $N$. Then, the above computational complexities become

$$c_{\text{FP}} = KB^2 + K^3 B \quad \text{and} \quad c_{\text{WMMSE}} = K^2 B^2, \tag{59}$$

so Algorithm 2 is more complex if the number of users $K$ is large. However, as already mentioned in Section V-B, because the matching weights are in practice expressed with finite precision, the efficiency of bipartite matching can be improved from $O(K^3)$ to $O(K^2)$ by using the algorithm of [34]. Then, we have

$$c_{\text{FP}} = KB^2 + K^2 B < K^2 B^2 = c_{\text{WMMSE}}. \tag{60}$$

In this case, Algorithm 2 is overall more computationally efficient than WMMSE.

## VII. CONCLUSION

This paper explores the application of FP for the discrete (or mixed discrete-continuous) problems for the communication system design. The central idea is to decouple the complicated interfering interactions among the different links by a novel quadratic transform and a Lagrangian dual transform, thereby allowing efficient and distributed optimization. This paper illustrates the proposed FP approach by considering the uplink user scheduling, power control, and beamforming problem for wireless cellular networks. By incorporating weighted bipartite matching, this paper devises a novel use of FP whereby the discrete scheduling variables can be jointly optimized with the continuous variables such as power and beamformers. As compared to the existing methods, the proposed FP approach treats discrete optimization rigorously without relaxation. The

$$\check{f}_q(\mathbf{s}, \mathbf{V}, \boldsymbol{\gamma}, \mathbf{Y}) = \sum_{(i,m)} w_{s_{im}} \left( \log(1 + \gamma_{im}) - \gamma_{im} + (1 + \gamma_{im}) \left( 2\text{Re}\left\{ \mathbf{v}_{s_{im}}^\dagger \mathbf{H}_{i,s_{im}}^\dagger \mathbf{y}_{im} \right\} - \mathbf{y}_{im}^\dagger \left( \sigma^2 \mathbf{I} + \sum_{(j,n)} \mathbf{H}_{i,s_{jn}} \mathbf{v}_{s_{jn}} \mathbf{v}_{s_{jn}}^\dagger \mathbf{H}_{i,s_{jn}}^\dagger \right) \mathbf{y}_{im} \right) \right) \tag{56}$$

paper further shows that the well-known WMMSE algorithm is a particular form of FP, but in contrast to the proposed approach, WMMSE is not well equipped to deal with discrete user scheduling variables. As a final remark, we mention that many other discrete optimization problems in communication system design are closely related to scheduling. Thus, the proposed FP approach can have wider applications in communication system design.

## APPENDIX A
## PROOFS OF PROPOSITIONS 1 AND 2

We focuses on proving Proposition 2, as Proposition 1 is just the scalar case of Proposition 2. Moreover, the convergence of the closed-form FP approach presented in Part I can also be established using Proposition 2 (by assuming fixed discrete variables). First, we introduce two useful lemmas, which can be easily verified.

*Lemma 1:* $f_o(\mathbf{s}, \mathbf{V}) \geq f_r(\mathbf{s}, \mathbf{V}, \boldsymbol{\gamma})$, with equality if and only if $\boldsymbol{\gamma}$ satisfies (38).

*Lemma 2:* $f_r(\mathbf{s}, \mathbf{V}, \boldsymbol{\gamma}) \geq f_q(\mathbf{s}, \mathbf{V}, \boldsymbol{\gamma}, \mathbf{Y})$, with equality if and only if $\mathbf{Y}$ satisfies (41).

Introduce a superscript $t$ to each variable as the iteration index in Algorithm 2, e.g., $\mathbf{V}^{(t)}$ refers to the set of transmit beamformers at the end of the $t$-th iteration. The auxiliary variable $\boldsymbol{\gamma}^{(t)}$ is determined by (38) using $(\mathbf{s}^{(t)}, \mathbf{V}^{(t)})$; similarly $\mathbf{Y}^{(t)}$ is determined by (41) using $(\mathbf{s}^{(t)}, \mathbf{V}^{(t)}, \boldsymbol{\gamma}^{(t)})$. But, define $\widetilde{\mathbf{Y}}^{(t)}$ to be the result of (41) using $(\mathbf{s}^{(t+1)}, \mathbf{V}^{(t+1)}, \boldsymbol{\gamma}^{(t)})$. It can be shown that:

$$
\begin{aligned}
f_o\left(\mathbf{s}^{(t+1)}, \mathbf{V}^{(t+1)}\right) &\stackrel{(a)}{=} f_r\left(\mathbf{s}^{(t+1)}, \mathbf{V}^{(t+1)}, \boldsymbol{\gamma}^{(t+1)}\right) \\
&\stackrel{(b)}{\geq} f_r\left(\mathbf{s}^{(t+1)}, \mathbf{V}^{(t+1)}, \boldsymbol{\gamma}^{(t)}\right) \\
&\stackrel{(c)}{=} f_q\left(\mathbf{s}^{(t+1)}, \mathbf{V}^{(t+1)}, \boldsymbol{\gamma}^{(t)}, \widetilde{\mathbf{Y}}^{(t)}\right) \\
&\stackrel{(d)}{\geq} f_q\left(\mathbf{s}^{(t+1)}, \mathbf{V}^{(t+1)}, \boldsymbol{\gamma}^{(t)}, \mathbf{Y}^{(t)}\right) \\
&\stackrel{(e)}{\geq} f_q\left(\mathbf{s}^{(t)}, \mathbf{V}^{(t)}, \boldsymbol{\gamma}^{(t)}, \mathbf{Y}^{(t)}\right) \\
&\stackrel{(f)}{=} f_r\left(\mathbf{s}^{(t)}, \mathbf{V}^{(t)}, \boldsymbol{\gamma}^{(t)}\right) \\
&\stackrel{(g)}{=} f_o\left(\mathbf{s}^{(t)}, \mathbf{V}^{(t)}\right)
\end{aligned}
$$

where $(a)$ follows by Lemma 1; $(b)$ follows since the update of $\boldsymbol{\gamma}$ in (38) maximizes $f_r$ when the other variables are fixed; $(c)$ follows by Lemma 2; $(d)$ follows since the update of $\mathbf{Y}$ in (41) maximizes $f_q$ when the other variables are fixed; $(e)$ follows since the joint updates of $\mathbf{s}$ and $\mathbf{V}$ in (46) and (47) maximize $f_q$ when the other variables are fixed; $(f)$ follows by Lemma 2; $(g)$ follows by Lemma 1.

Therefore, the weighted sum rate objective $f_o$ is monotonically nondecreasing after each iteration. Since the value of $f_o$ is bounded above, the algorithm must converge. At the convergence, the algorithm arrives at a local optimum of the reformulated problem of $f_q$. Further, for fixed scheduling variable $\mathbf{s}$, the solution is a stationary point of the original problem of $f_o$ in $\mathbf{V}$.

**Kaiming Shen** (S'13) received the B.Eng. degree in information security and the B.S. degree in mathematics both from Shanghai Jiao Tong University, Shanghai, China in 2011, and the M.A.Sc. degree in electrical and computer engineering from the University of Toronto, Ontario, Canada in 2013. He is currently pursuing the Ph.D. degree at the University of Toronto.

His research interests include mathematical optimization, algorithms for wireless networks, and information theory.

**Wei Yu** (S'97-M'02-SM'08-F14) received the B.A.Sc. degree in Computer Engineering and Mathematics from the University of Waterloo, Waterloo, Ontario, Canada in 1997 and M.S. and Ph.D. degrees in Electrical Engineering from Stanford University, Stanford, CA, in 1998 and 2002, respectively. Since 2002, he has been with the Electrical and Computer Engineering Department at the University of Toronto, Toronto, Ontario, Canada, where he is now Professor and holds a Canada Research Chair (Tier 1) in Information Theory and Wireless Communications. His main research interests include information theory, optimization, wireless communications and broadband access networks.

Prof. Wei Yu currently serves on the IEEE Information Theory Society Board of Governors (2015-20). He was an IEEE Communications Society Distinguished Lecturer (2015-16). He is currently an Area Editor for the IEEE Transactions on Wireless Communications (2017-20). He served as an Associate Editor for IEEE Transactions on Information Theory (2010-2013), as an Editor for IEEE Transactions on Communications (2009-2011), and as an Editor for IEEE Transactions on Wireless Communications (2004-2007). He is currently the Chair of the Signal Processing for Communications and Networking Technical Committee of the IEEE Signal Processing Society (2017-18) and served as a member in 2008-2013. Prof. Wei Yu received the Steacie Memorial Fellowship in 2015, the IEEE Signal Processing Society Best Paper Award in 2017 and 2008, a Journal of Communications and Networks Best Paper Award in 2017, an IEEE Communications Society Best Tutorial Paper Award in 2015, an IEEE ICC Best Paper Award in 2013, the McCharles Prize for Early Career Research Distinction in 2008, the Early Career Teaching Award from the Faculty of Applied Science and Engineering, University of Toronto in 2007, and an Early Researcher Award from Ontario in 2006. Prof. Wei Yu is a Fellow of the Canadian Academy of Engineering, and a member of the College of New Scholars, Artists and Scientists of the Royal Society of Canada. He is recognized as a Highly Cited Researcher.